\documentclass[fleqn,usenatbib]{mnras}

\usepackage{newtxtext,newtxmath}

\usepackage[T1]{fontenc}

\DeclareRobustCommand{\VAN}[3]{#2}
\let\VANthebibliography\thebibliography
\def\thebibliography{\DeclareRobustCommand{\VAN}[3]{##3}\VANthebibliography}

\usepackage{graphicx}	
\usepackage{amsmath}	
\usepackage{pdflscape}
\usepackage[dvipsnames]{xcolor}

\title[A possible extreme dwarf nova with year-long outbursts]{OGLE-BLG504.12.201843: A possible extreme dwarf nova \thanks{Based in parts on observations obtained at the Southern Astrophysical Research (SOAR) telescope, which is a joint project of the Minist\'{e}rio da Ci\^{e}ncia, Tecnologia e Inova\c{c}\~{o}es (MCTI/LNA) do Brasil, the US National Science Foundation’s NOIRLab, the University of North Carolina at Chapel Hill (UNC), and Michigan State University (MSU).}}

\author[C. Landri et al.]{Camille Landri,$^{1}$ \thanks{E-mail: \href{mailto:camille.landri@utf.mff.cuni.cz}{camille.landri@utf.mff.cuni.cz}} 
Ondřej Pejcha,$^{1}$  Micha\l{} Pawlak,$^{2}$ Andrzej Udalski,$^{3}$ Jose L. Prieto,$^{4,5}$
\newauthor Manuel Barrientos,$^{6}$ Jay Strader$^{7}$ and Subo Dong$^{8}$
\\
$^{1}$Institute of Theoretical Physics, Faculty of Mathematics and Physics, Charles University, V Holešovičkách 2, 180 00 Praha 8, Czech Republic\\
$^{2}$Astronomical Observatory, Jagiellonian University, ul. Orla 171, 30-244 Kraków, Poland\\
$^{3}$Astronomical Observatory, University of Warsaw, Al. Ujazdowskie
4, 00-478 Warszawa, Poland\\
$^{4}$Núcleo de Astronomía de la Facultad de Ingeniería y Ciencias, Universidad Diego Portales, Av. Ejército 441, Santiago, Chile\\
$^{5}$Millennium Institute of Astrophysics, Santiago, Chile\\
$^{6}$Homer L. Dodge Department of Physics and Astronomy, University of Oklahoma, 440 W. Brooks St., Norman, OK, 73019 USA\\
$^{7}$Department of Physics and Astronomy, Michigan State University, East Lansing, MI 48824\\
$^{8}$Kavli Institute for Astronomy and Astrophysics, Peking University, Yi He Yuan Road 5, Hai Dian District, Beijing 100871, China\\
}

\date{Accepted 2022 September 29. Received 2022 September 23; in original form 2022 January 31}

\pubyear{2022}

\begin{document}
\label{firstpage}
\pagerange{\pageref{firstpage}--\pageref{lastpage}}
\maketitle

\begin{abstract}
We present the analysis of existing optical photometry and new optical spectroscopy of the candidate cataclysmic variable star OGLE-BLG504.12.201843. As was shown previously, this object has an orbital period of 0.523419~days and exhibits year-long outbursts with a mean period of 973~days. Using digitized photographic archives, we show that the earliest recorded outburst occurred in 1910. We propose that this object is a U Gem-type dwarf nova with extreme properties. The orbital variability of the system in outburst shows clear signs of an accretion disc, from which the outburst likely originates. During quiescence, the object slowly brightens by up to $0.75$~mag in the $I$~band over 600~days before the outburst and exhibits small flares with amplitude $\lesssim 0.2$~mag in the $I$~band. We interpret the gradual brightening as an increase in the luminosity and temperature of the accretion disc, which is theoretically predicted but only rarely seen in DNe. The origin of small flares remains unexplained. The spectra shows Balmer absorption lines both in quiescence and outburst, which can be associated with a bright secondary star or a cold accretion disc. During outbursts, emission lines with FWHM of about 450~km s$^{-1}$ appear, but they lack typical double-peaked profiles. We suggest that either these lines originate in the disc winds or the orbital inclination is low, the latter being consistent with constrains obtained from the orbital variability of the system. Due to its extreme properties and peculiarities, OGLE-BLG504.12.201843 is an excellent object for further follow-up studies.

\end{abstract}

\begin{keywords}
stars: dwarf novae -- accretion discs -- binaries: close -- novae, cataclysmic variables
\end{keywords}



\section{Introduction}
OGLE-BLG504.12.201843 (hereafter O-201843) is a candidate cataclysmic variable discovered by \cite{Mroz.2016} with the \textit{Optical Gravitational Lensing Experiment} (OGLE, \citealt{Udalski.2015}) in the field BLG504.12 and is located at RA=17:57:19.65, Dec.=-28:08:15.7 (J2000). \cite{Mroz.2016} found that O-201843 undergoes 300-days long outbursts that repeat every 950 to 1020~days. The outburst amplitude is $1.75$~mag in the $I$~band. \cite{Mroz.2016} also detected a photometric variability with a period of 0.523419~days, which they interpret as the orbital period. In quiescence, the orbital light curve has a double-hump profile, but as the system gets brighter the minima change shape and depth. At maximum brightness, the orbital light curve shows a single hump. Additionally, \cite{Mroz.2016} noticed an initially very slow brightening to the maximum, which abruptly accelerates. They speculated that the long cycles are caused by interactions with a tertiary star, but no sign of tertiary orbit was detected in the O-C analysis. Overall the photometry of the system is reminiscent of dwarf novae (DN), albeit with extreme properties. Motivated by these intriguing features and especially the slow rise from the quiescence, we selected O-201843 for more detailed analysis. If O-201843 indeed is a DNe, studying its extreme properties could lead to a better understanding of the mechanisms responsible for DN outbursts.

DNe are a type of CVs that undergo semi-periodic outbursts \citep{warner.2003}. These systems are composed of a White Dwarf (WD) accreting matter from a secondary star, typically located on the main sequence. The accreting material forms a disc, which can become thermally unstable and develop outbursts. Some DNe only undergo regular 2-5~mag outbursts (U Gem type) while others display additional features. Examples of these features are longer and less recurrent outbursts called superoutbursts (SU~UMa type) or standstills that interrupt a sequence of outbursts (Z Cam type).  

The outbursts of DNe are currently best explained by the Disc Instability Model \citep[DIM, e.g.][]{meyer.1981,smak.1982,cannizzo.1982,faulkner.1983,mineshige.1983}. This model is based on a Shakura-Sunyaev disc \citep{Shakura.1973} supplemented with additional physical processes, such as mass transfer variations, inner disc truncation, disc winds and irradiation of the disc, which have been reviewed by \cite{Lasota.2001} and \cite{hameury.2020}. For such a disc, the effective temperature as a function of the surface density (S-curve) describes cold and a hot stable branches and a viscously and thermally unstable region between them. 

In DNe, the disc has local mass transfer rates that lie in the instability range. During quiescence, matter accumulates until the mass transfer reaches the upper critical value and heat fronts propagate both inwards and outwards. The temperature rises until the peak of the outburst, then a cooling front starts propagating from the outside of the disc and the disc goes back to quiescence. The radius at which the heating front starts has an impact on the duration and shape of the outburst. If it starts far from the inner edge of the disc (i.e. the disc is almost stable), the outburst is asymmetrical: the decline is slow at first and then accelerates to reach values similar to the rise. This outside-in outburst involves a large part of the mass of disc. Alternatively, the heating front can develop close to the inner edge of the disc, which leads to more symmetric outburst shapes. One prediction of the DIM is the critical mass transfer rate that separates DNe and Nova-like CVs which seems to match the observations \citep{dubus.2018}.

DN outbursts are never strictly identical because there are differences in parameters that control the outburst, such as the radius at which the outburst is triggered, the distribution of material after the last outburst, among others. However one can find correlations between some of the characteristics of the outbursts. The two most probable relations are the Kukarkin-Parenago relation between the amplitude and the average recurrence time of outbursts \citep{kukarkin.1934} and the Bailey relation between the rate of brightness decay and the orbital period \citep{Bailey.1975}. Recently, \cite{otulakowska.2016} presented a statistical analysis of the measurable properties of a large sample of DNe outbursts in order to find possible correlations between the different characteristics. They were not able to verify the Kukarkin-Parenago relation but confirmed the Bailey relation as well as other connections, e.g. the correlation between outburst duration and orbital period.

In this paper, we aim to better constrain the nature of O-201843 by analyzing its optical photometry and spectroscopy. In Section~\ref{sec:observations}, we describe our photometric and spectroscopic dataset. In Section~\ref{sec:results}, we discuss the different features of the system. In Section~\ref{sec:discussion}, we speculate about the origins of these features, compare O-201843 to DNe and try to find similarities that would allow to characterise the system. In Section~\ref{sec:conclusion}, we summarize our results.

\section{Observations} \label{sec:observations}

\subsection{Photometry}

\begin{figure*}
    \includegraphics[width=0.95\linewidth]{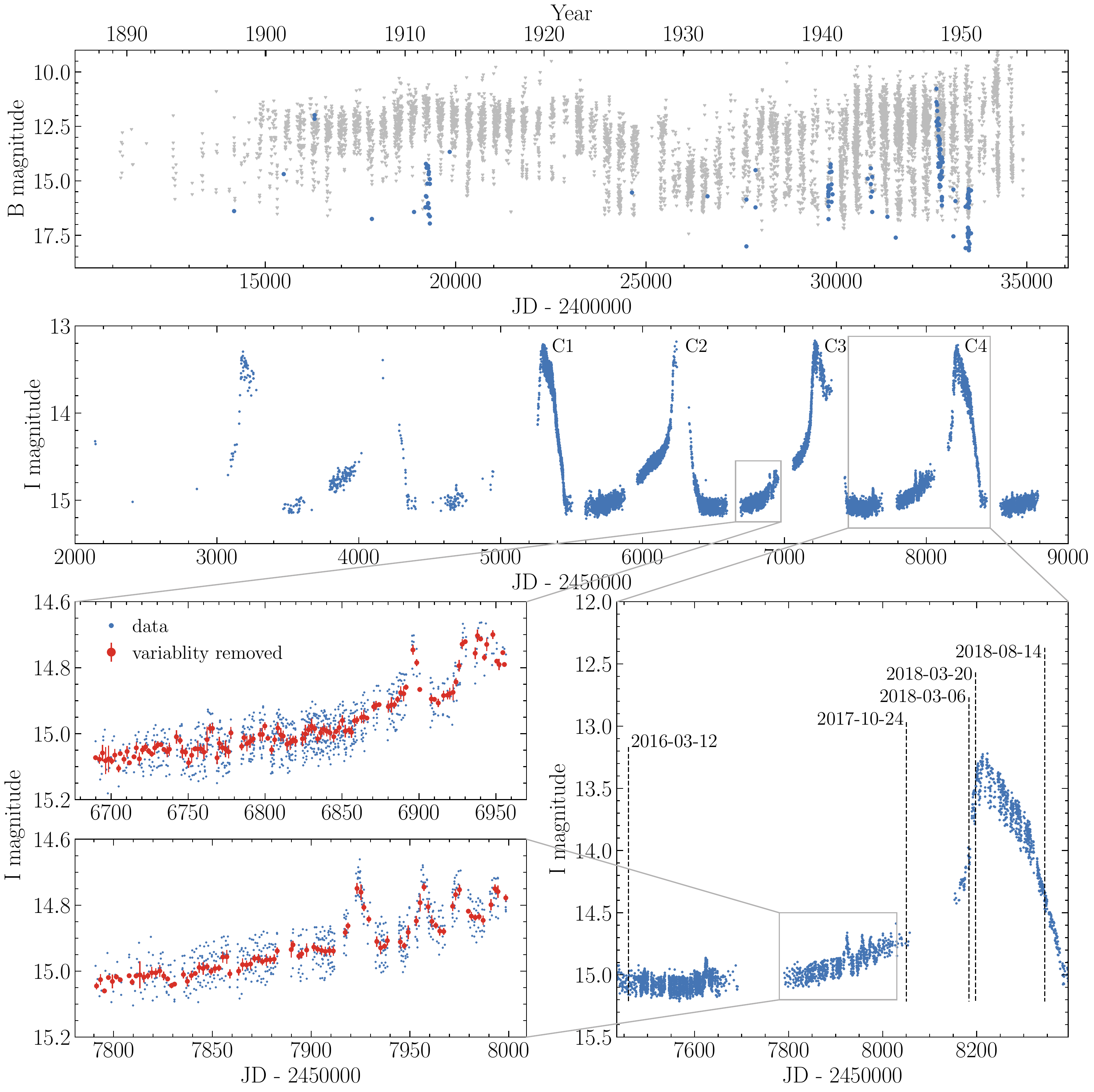}
    \caption{Photometry of OGLE-BLG504.12.201843. The first panel shows data from DASCH, with detections in blue and non-detections in grey. The middle panel shows data from OGLE, we denote the cycles of interest as C1, C2, C3 and C4. The right panel shows one outburst cycle with dates at which spectra were obtained, which correspond to three out of the four phases of the light curve. The left panels show the small flares appearing in OGLE data in more details. The data is shown in blue and the phase-averaged brightness is shown in red. The latter was obtained by subtracting a fit of the variability (made with with Fourier series) from the data and averaging the result over bins of 2~days (in red).}
    \label{fig:photometry}
\end{figure*}

In Figure~\ref{fig:photometry}, we show available optical time-series measurements of O-201843. We use $I$~band observations from OGLE, which cover 2001--2021. The data up to year 2015 were published by \cite{Mroz.2016}, the later data are presented here for the first time. As seen in Figure~\ref{fig:photometry}, the system was irregularly observed between 2001 and 2009 and then monitored from 2010 to 2019 at a roughly regular 1-day cadence with several interruptions. A total of six bright outbursts were recorded, with a recurrence timescale of roughly 1000~days. The system has a brightness of  $I=15$~mag in quiescence and peaks at $I=13.25$~mag during the outbursts; we discuss the outburst properties in Section~\ref{sec:outbursts}. From the photometric variability of the system, \cite{Mroz.2016} detected an orbital period of 0.523419~days, we discuss the orbital variability in Section~\ref{sec:variability}. We see in the left panel of Figure~\ref{fig:photometry} that small $\Delta I=0.2$~mag flares appear during quiescence, a feature that was not discussed by \cite{Mroz.2016}. We provide more details on the flares in Section~\ref{sec:flares}.

Additionally, we looked for archival data from the Digital Access to a Sky Century @ Harvard (DASCH) catalog \citep{laylock.2010}. The DASCH project digitizes plates from the Astronomical Photographic Plate Collection and converts them to photometry, allowing the study of the sky on 100~years timescales. Observations from photographic plates roughly correspond to $B$~band photometric measurements. We found a total of 154 detections made between 1903 and 1951 which are shown in the top panel of Figure~\ref{fig:photometry} along with the non-detections. The system displays five distinguishable peaks in 1911, 1940, 1943, 1948 and 1950, showing that this system has gone through these outbursts in the past. The time elapsed between the outbursts seems to roughly match the periodicity of the recent observations. The peak magnitude of the outburst averages around 14~mag but one peak goes up to 11~mag, however, we do not know if this peak was extraordinary or if the other peaks were only partly recorded.

\subsection{Spectroscopy}

We obtained six spectra of O-201843 between March~12~2016 and August~14~2018. To our knowledge, these are the first spectra of O-201843. Four spectra are Echelle spectra and were taken with the MIKE Spectrograph on the Clay telescope at the Las Companas Observatory (LCO). The two others were obtained with the Goodman High Throughput Spectrograph at the Southern Astrophysical Research Telescope (SOAR) (see Table~\ref{tab:spectralog}). The bottom right panel of Figure~\ref{fig:photometry} shows when the spectra were taken in the context of the outbursts. Two spectra were taken during the quiescence of the light curve, three during the rise and the last one during the decline. The data were reduced using IRAF and after removal of the continuum we performed Gaussian fits on the well-defined lines. The results are listed in Table~\ref{tab:spectra} and the evolution of the most prominent feature is shown in Figure~\ref{fig:HeII}. Larger portions of our spectra with lines identified in Table~\ref{tab:spectra} are shown in Appendix~\ref{sec:spe}. 

\begin{figure}
    \centering
    \includegraphics[width=0.8\columnwidth]{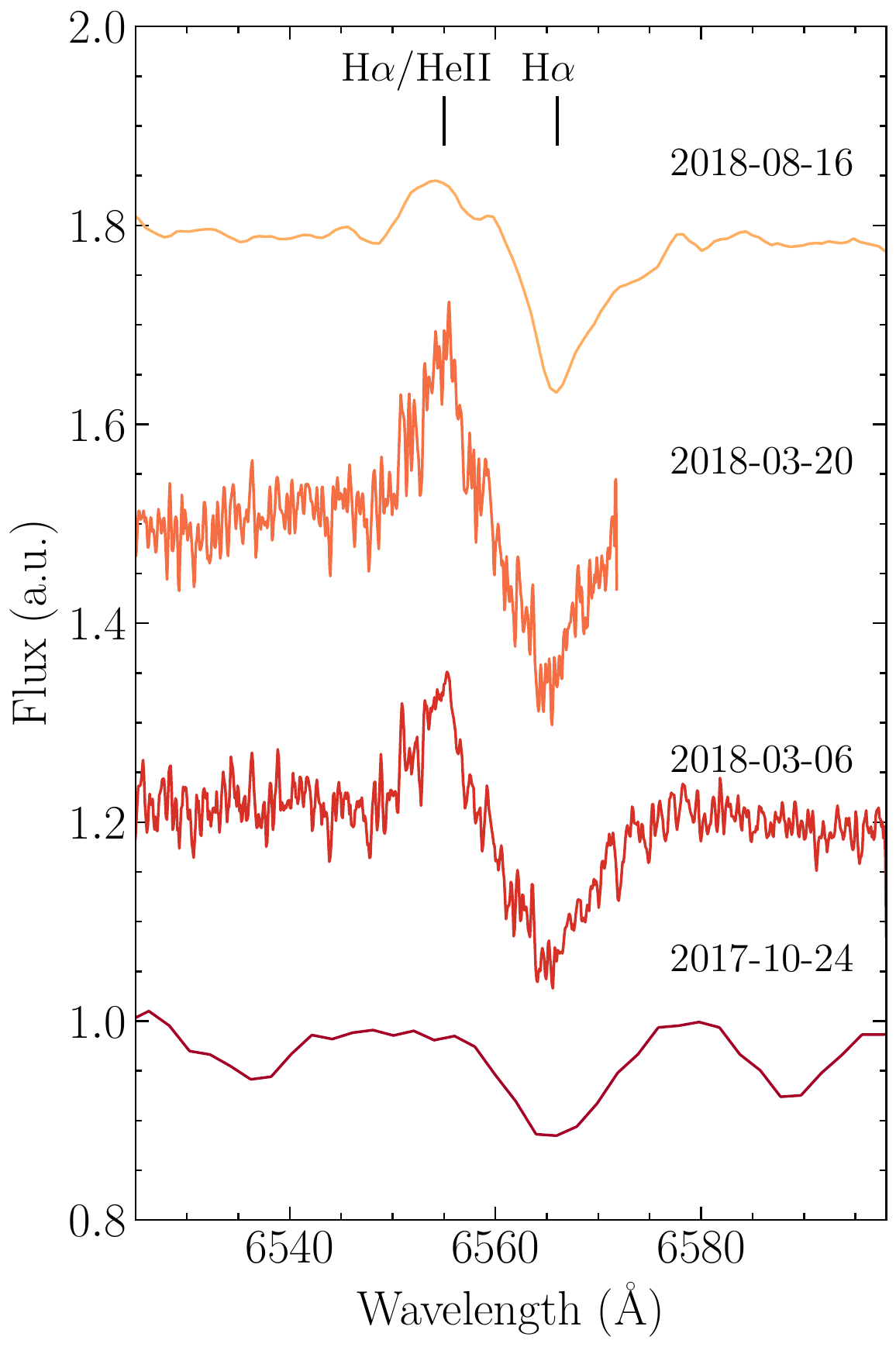}
    \caption{Evolution of the H$\alpha$ absorption line and the 6560~Å emission line at different epochs. The spectrum taken on the 2017-10-24 corresponds to the quiescence of the system while the three other spectra were taken during an outburst.}
    \label{fig:HeII}
\end{figure}

\begingroup
\begin{table*}
\caption {\label{tab:spectralog}Log of spectroscopic observations of O-201843.} 
\begin{tabular}{llllllll}
\hline
\multicolumn{2}{c}{Time of observation} & telescope & instrument & wavelength coverage & spectral resolution & observer & reduction \\ 
\cline{1-2} 
Date   & Time (UT)     & & & (Å) & & &                                                \\    \hline
2016-03-12 & 10:08 & Magellan & IMACS & 3650-9740 & & Dong, S & Prieto, JL \\
2017-10-24 & 00:22:03 & SOAR 4.1m & Goodman Spectrograph & 3200-9000 & 1850 & Strader, J & Strader, J\\
2017-10-24 & 00:22:03 & SOAR 4.1m & Goodman Spectrograph & 3200-9000 & 1850 & Strader, J & Strader, J\\
2018-03-06 & 13:52:22 & Magellan & MIKE-Blue & 3350-5000 & 83000 & Barrientos, M. & Prieto, JL \\
2018-03-06 & 13:52:25 & Magellan & MIKE-Red & 4900-9500 & 65000 & Barrientos, M. & Prieto, JL \\
2018-03-20 & 13:51:00 & Magellan & MIKE-Red & 4900-9500 & 65000 & Barrientos, M. & Prieto, JL \\
2018-08-14 & 02:41:27 & SOAR 4.1m & Goodman Spectrograph & 3200-9000 & 5880 & Strader, J & Strader, J\\
\hline
\end{tabular}
\end{table*}
\endgroup

\begingroup
\begin{table*}
\caption {\label{tab:spectra}Most prominent spectral lines of O-201843 and their evolution. The parameters of each lines are obtained by fitting a Gaussian profile.} 
\begin{tabular}{lllllll}
\hline
Date & Wavelength & Line & Type & \multicolumn{2}{c}{FWHM} & Eq. Width  \\ 
\cline{5-6}
      &  (Å)      &      &      & (Å)  & (km/s)            &  (Å)   \\    \hline
2016-03-12  & 6557.8 & H$\alpha$ & Absorption & 10.61 $\pm$ 1.25 & 485 $\pm$ 57 & 3.77 $\pm$ 0.45 \\ \hline
2017-10-24  & 4342.7 & H$\gamma$ & Absorption & 7.90 $\pm$ 1.31 & 545 $\pm$ 90 & 2.41 $\pm$ 0.38 \\
            & 6565.9 & H$\alpha$ & Absorption & 9.63 $\pm$ 1.63 & 440 $\pm$ 74 & 1.11 $\pm$ 0.18 \\ \hline
2018-03-06  & 4341.3 & H$\gamma$ & Absorption & 8.05 $\pm$ 0.19 & 556 $\pm$ 13 & 1.52 $\pm$ 0.04 \\
            & 4680.3 & HeII - 4686 & Emission & 7.02 $\pm$ 0.82 & 450 $\pm$ 53 & 1.21 $\pm$ 0.19 \\
            & 4862.8 & H$\beta$ & Absorption & 7.56 $\pm$ 0.19 & 466 $\pm$ 11 & 1.32 $\pm$ 0.03 \\
            & 6554.2 & HeII - 6560 or H$\alpha$ & Emission & 5.21 $\pm$ 0.89 & 238 $\pm$ 41 & 0.71 $\pm$ 0.11 \\
            & 6565.8 & H$\alpha$ & Absorption & 8.47 $\pm$ 0.61 & 387 $\pm$ 28 & 1.37 $\pm$ 0.10 \\ \hline
2018-03-20  & 6554.2 & HeII - 6560 or H$\alpha$ & Emission & 6.34 $\pm$ 0.42 & 290 $\pm$ 19 & 0.94 $\pm$ 0.06 \\
            & 6565.5 & H$\alpha$ & Absorption & 7.24 $\pm$ 0.33 & 331 $\pm$ 15 & 1.07 $\pm$ 0.05 \\ \hline
2018-08-14  & 6554.1 & HeII - 6560 or H$\alpha$ & Emission & 5.81 $\pm$ 1.7 & 266 $\pm$ 77 & 0.37 $\pm$ 0.09 \\
            & 6567.2 & H$\alpha$ & Absorption & 8.59 $\pm$ 0.4 & 392 $\pm$ 18 & 1.36 $\pm$ 0.06 \\ \hline
\end{tabular}
\end{table*}
\endgroup

\section{Results} \label{sec:results}
In this Section, we examine the photometric and spectroscopic features of O-201843. We discuss the outburst shapes and periodicity in Section~\ref{sec:outbursts} and study the evolution of the phased light curves during the outbursts in Section~\ref{sec:variability}. In Section~\ref{sec:flares}, we look at the newly identified flares. Finally we investigate the spectroscopy and its evolution during the outbursts in Section~\ref{sec:spectranalysis}.

\subsection{Outbursts}\label{sec:outbursts}
We first investigate the general shape of the outburst cycles. The middle panel of Figure \ref{fig:photometry} shows the six outbursts that were recorded during the 18~years of observation. They are not strictly periodic but happen approximately every 1000 days and they display a 1.75~mag increase in the $I$~band. Looking at the right panel of Figure~\ref{fig:photometry}, we estimate the outburst duration to be around 300~days. We also see an intermediate state where the luminosity slowly increases by about 0.75~mag before the outburst starts, i.e. almost half of the total luminosity increase of the system. We choose to consider it as a part of the quiescence of the system, which makes the low state 700~days long. In order to investigate the periodicity of the outbursts we performed an $O-C$ analysis of the light curve. The maxima of outbursts were obtained using spline fits of the light curve. We improved the average period obtained from this analysis by correcting it with the change in period and reapplying the $O-C$ fit, until this change becomes negligible. This yields an average recurrence time of $P_{\text{cycle}}=973.16$~days. We checked the timings of maxima from OGLE data and we did not find any significant trend. Unfortunately, the data from DASCH are too sparse to allow meaningful fits of maxima.

\begin{figure}
    \includegraphics[width=\columnwidth]{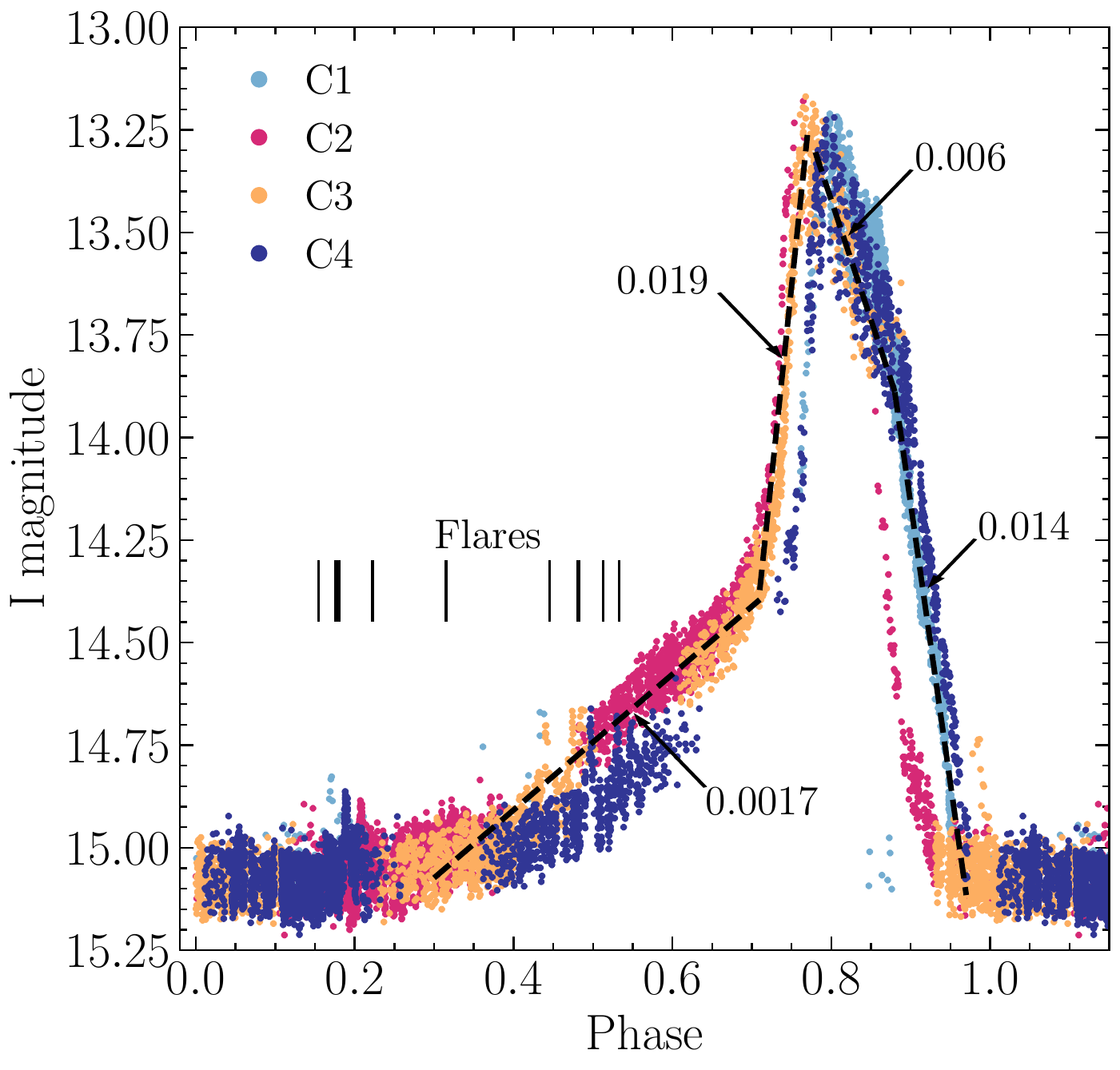}
    \caption{Light curve folded over the recurrence of the outburst (973.16~days) using Equation~\eqref{eq:cyclefold}. We show four cycles which are each denoted by a different colour. We indicate the time at which the flares discussed in Section~\ref{sec:flares} were recorded with black vertical lines. The average slopes of the different phases of the cycle are plotted in black with the values (in mag.day$^{-1}$) indicated next to each phase.}
    \label{fig:outburstrecurrence}
\end{figure}

To check if the shape of the outburst evolves, we fold the light curve over the recurrence time using
\begin{equation}\label{eq:cyclefold}
 JD = 2451601.62 + 973.16 \times E,
\end{equation}
where $E$ is the epoch. We show the results in Figure~\ref{fig:outburstrecurrence} and we see that the shape of the outburst does not change significantly from one cycle to another. This is emphasised by the consistence of the rise and decline rates throughout the cycles. In fact, the outbursts show slow and fast decline rates that remain around 0.006 and 0.014~mag/day, while the quiescence and outburst rise vary between 0.0017 to 0.0019~mag/day and 0.016 to 0.021~mag/day, respectively.

\begin{figure*}
    \includegraphics[width=1.\linewidth]{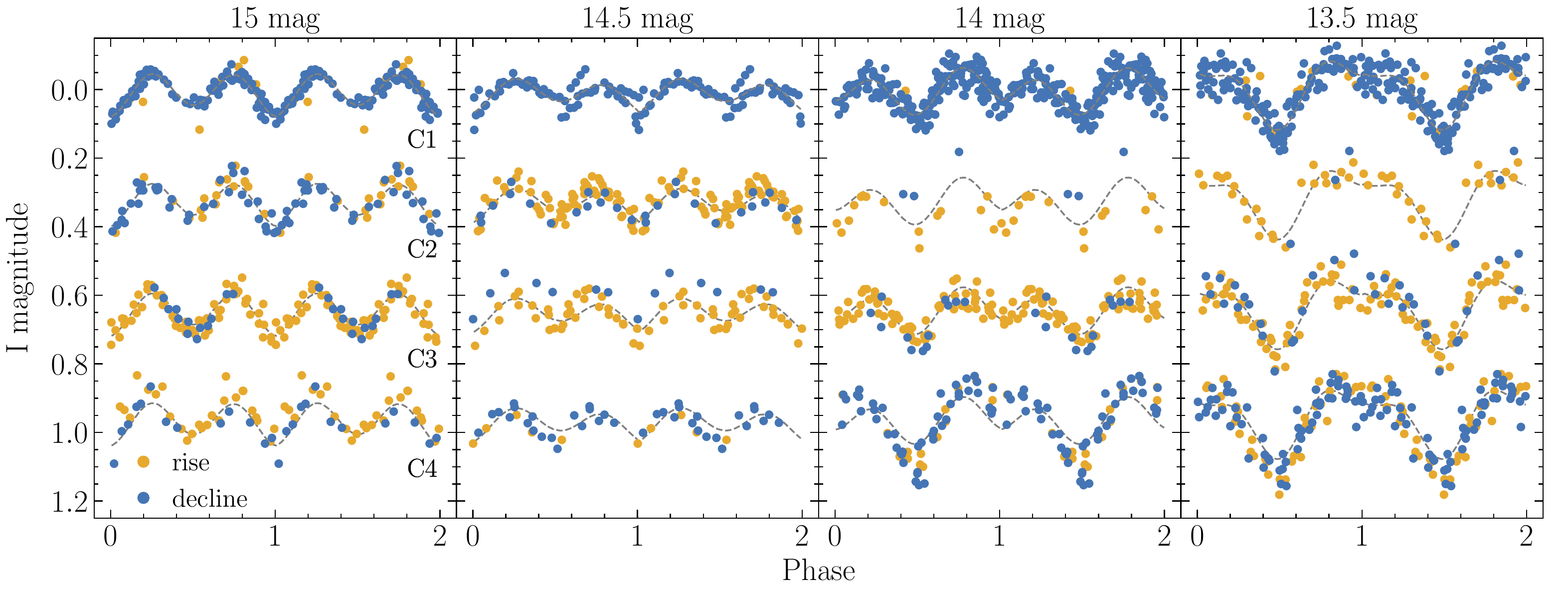}
    \caption{Photometry of OGLE-BLG504.12.201843 during cycles C1, C2, C3 and C4 folded around its orbital period (0.523419 days) using Equation~\eqref{eq:orbfold}. We separated the data in 4 magnitude bins (15, 14.5, 14 and 13.5~mag) and removed the effect of the outburst on the photometry to single out the orbital variations of the system. We also separate each bin in ``rise'' and ``decline'' phases of the outburst, shown in yellow and blue respectively. For each bin, we plot a fit of the cycle with the clearest shape in grey for comparison with the other cycles.}
    \label{fig:photometryphase}
\end{figure*}

\subsection{Orbital variability}\label{sec:variability}
Following the orbital period found by \cite{Mroz.2016} (0.523419~day), we show the evolution of the orbital variability of the system in Figure~\ref{fig:photometryphase}. This was obtained by fitting the general trend of the light curve using splines and subtracting the fit from the data. The result is folded over the orbital period following
\begin{equation} \label{eq:orbfold}
 JD = 2452141.23 + 0.523419 \times E,
\end{equation}
and is repeated over two periods for readability purposes. We show the variability of the light curve for 20 orbital periods around the time at which a specific phase of the outburst is reached. These phases correspond to different values of brightness, i.e 13.5~mag is the peak of the outburst, 15~mag is the low quiescence, 14 and 14.5~mag are intermediate states. The data are also split in rising and declining phases of the light curve in order to evaluate differences in shape. We overplotted a fit of the cycle with the most data for comparison purposes.

As shown by \cite{Mroz.2016}, the variability shape is different during quiescence and outburst. During quiescence, an orbital period shows two maxima and two minima due to the tidal distortion of the secondary. Additionally, the two minima of the light curve are not equal, which can be caused by different mechanisms. A possible explanation is the irradiation of the secondary by the primary, which brightens the side of the secondary that faces the primary. An other possibility is gravity darkening: the distortion of the secondary lowers the surface gravity on the side that faces the primary, changing the temperature and pressure conditions for hydrostatic equilibrium and darkening of the area \citep{lara.2012}. This results in the dimming of the side of the secondary that is facing the primary. These ellipsoidal modulations also mean that the secondary star contributes significantly to the observed flux.

During the outburst, the minimum at phase~0.5 becomes deeper and sharper while the minimum at phase~0 is absorbed by the brightness of the outburst. The deeper minima is likely caused by the secondary eclipsing the component of the system from which the outburst originates. The eclipse at phase~0.5 means that the corresponding minimum in quiescence occurs when the back of the secondary is observable and the deeper minimum at phase~0 occurs when we observe the side facing the primary. Therefore, the difference in minima during quiescence seems to be caused by gravity darkening rather than irradiation.

We note that during the second part of the outburst decay, the shape of the variability has already switched back to the quiet state, so the outbursting component is not luminous enough to hide the tidal distortion of the secondary. We also look for changes in the variability between the rise and the decline as well as between different outbursts. We do not find significant differences and conclude that the brightness contribution of the different components of the system does not change from one outburst to the other or from rise to decline.

\subsection{Flares}\label{sec:flares}
Another peculiar feature of O-201843 are the small flares with amplitudes up to $\Delta I=0.2$~mag appearing during the quiescence which were not discussed by \cite{Mroz.2016}. They are often barely apparent due to their very small amplitude. The most distinct ones appear in the last outburst recorded, shown in the bottom right panel of Figure~\ref{fig:photometry}.

The flares last around ten days, and in some instances they recur every five to twenty days in a 100-days interval. To distinguish the flares from the orbital variability, we fitted the orbital variations using third order Fourier series and subtracted the fit from the light curve. We averaged the result over bins of two days to remove the scatter of the data around the fit. This method confirmed ten flares, which are indicated in Figure~\ref{fig:outburstrecurrence}, and some of the light curves are shown in detail in the bottom left panels of Figure~\ref{fig:photometry}. Some flares are too hard to unambiguously identify because the observations are too sparse and the flare amplitude is comparable to the variability of the system. Additionally, they seem to only occur during quiescence, which is unfortunately never fully covered due to occultation by the Sun. A significant amount of flares might have not been detected, but from the observations it seems the flares happen at any phase of the quiescence.

\subsection{Spectral analysis}\label{sec:spectranalysis}
For each spectrum of O-201843, we fitted the interesting lines of the system with simple Doppler broadening in order to loosely quantify their evolution, we show the results in Table~\ref{tab:spectra}.

During quiescence, the spectra are mostly featureless. They show a slightly blue continuum with broad Balmer absorption, and Na and Ca~II lines coming from the interstellar medium. From comparison with other spectral types (see Fig.~\ref{fig:fullspectra}) we see that the Balmer lines might come from a late~A-type or early~F-type star. We note that the Balmer lines are relatively shallow compared to those of stellar spectra. Although the photometry of the system shows two stars, we see no signatures of an other star, which could be hidden by the brightness of its companion. The contribution of the secondary to the brightness of the system is significant enough to cause ellipsoidal modulations in the photometry (see \ref{sec:variability}). We therefore expect the brighter star to be the secondary. Since it is not possible to conclude on the nature of the primary star, we assume the usual primary of a CV, a WD, hidden by the bright secondary.

The three spectra obtained during outburst are Echelle spectra, from which we are unable to recover the shape of the continuum. We can therefore only use spectral lines to estimate the behaviour of the system in outburst. In these spectra, we find the same Balmer absorption lines as the one observed in quiescence. Considering the differences in resolution of the different spectra, the FWHM and equivalent width of these lines do not appear to change significantly during the outburst. We also find a He~II emission line at 4680~Å and an emission line at 6555~Å that could correspond to either He~II or H$\alpha$. Due to the limited range of our spectra, we only detect the He~II line at one epoch. The evolution of the 6555~Å line presented in Figure~\ref{fig:HeII} shows that it is the strongest near the peak of the outburst and fades away during the decline.

\section{Discussion}\label{sec:discussion}
The origin of the outbursts of O-201843 is unclear and is examined in this Section. The light curve of O-201843 is consistent with the predictions of the DIM, i.e. a thermally unstable accretion disc, comparable to those of DNe.

The DIM predicts that the brightness of the system should rise slowly before the outburst is triggered. This is a consequence of the gradual increase of the local mass transfer rate and temperature before reaching the critical values for stability. We can also use the DIM to explain the evolution of the orbital modulations. During quiescence, the disc is very dim, and the variability detected is due to the secondary. Mass starts to build up in the disc, it slowly brightens and starts to hide the effects of the distortion of the secondary. The brighter fraction of the disc is small at first but its extent grows as matter accumulates, affecting the two minima of the variability differently. One becomes shallower as the disc outshines the secondary and the other becomes deeper when the secondary obscure a small part of the disc. This effect is best seen when the disc is at its brightest, at the peak of the outburst (see the right panel of Figure~\ref{fig:photometryphase}). The shallow eclipse indicates that the inclination of the system is moderate, a higher inclination would deepen the eclipse as seen in eclipsing systems like IP~Peg \citep{Bobinger.1999} or OY~Car \citep{Nicholson.2009}.

We note that if the outbursts can be explained by the DIM, then the high amplitude of the outbursts translates into a high critical disc mass transfer rate $\dot{M}_\textrm{crit}$. Since $\dot{M}_\textrm{crit}$ increases with the disc radius \citep{Smak.1983}, we expect the extent of the disc to be large, which is consistent with the large dimensions of the system. The timescales of the outbursts also suggests that the quiescent disc is likely cold. Since the disc is usually expected to be irradiated by the WD, the low temperature of the disc would mean that the WD is relatively cold as well.

While the DIM can account for most of the photometric features of O-201843, the spectroscopy of the system differs from the spectra of thermally unstable accretion disc. We discuss this issue in Section~\ref{sec:accdisc} and we constrain the system geometry in Section~\ref{sec:geom}. We consider the ensuing conditions on the mass transfer between the secondary and the primary in Section~\ref{sec:masstransfer}. Finally, in order to contextualise the nature of O-201843 we compare it with known DNe in Section~\ref{sec:comparison}.

\subsection{Spectra of unstable accretion discs}\label{sec:accdisc}

\begin{figure*}
    \includegraphics[width=0.75\columnwidth]{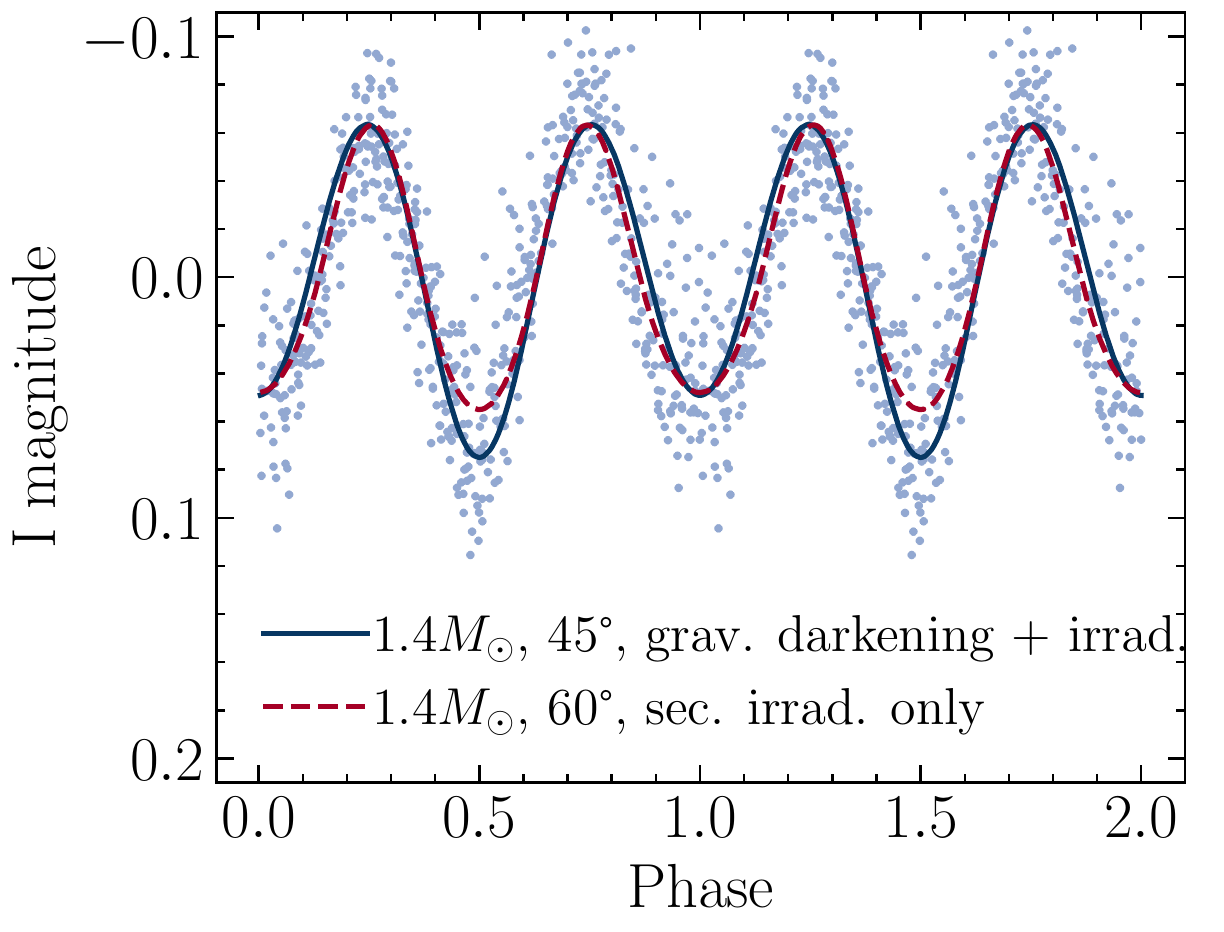}
    \includegraphics[width=0.75\columnwidth]{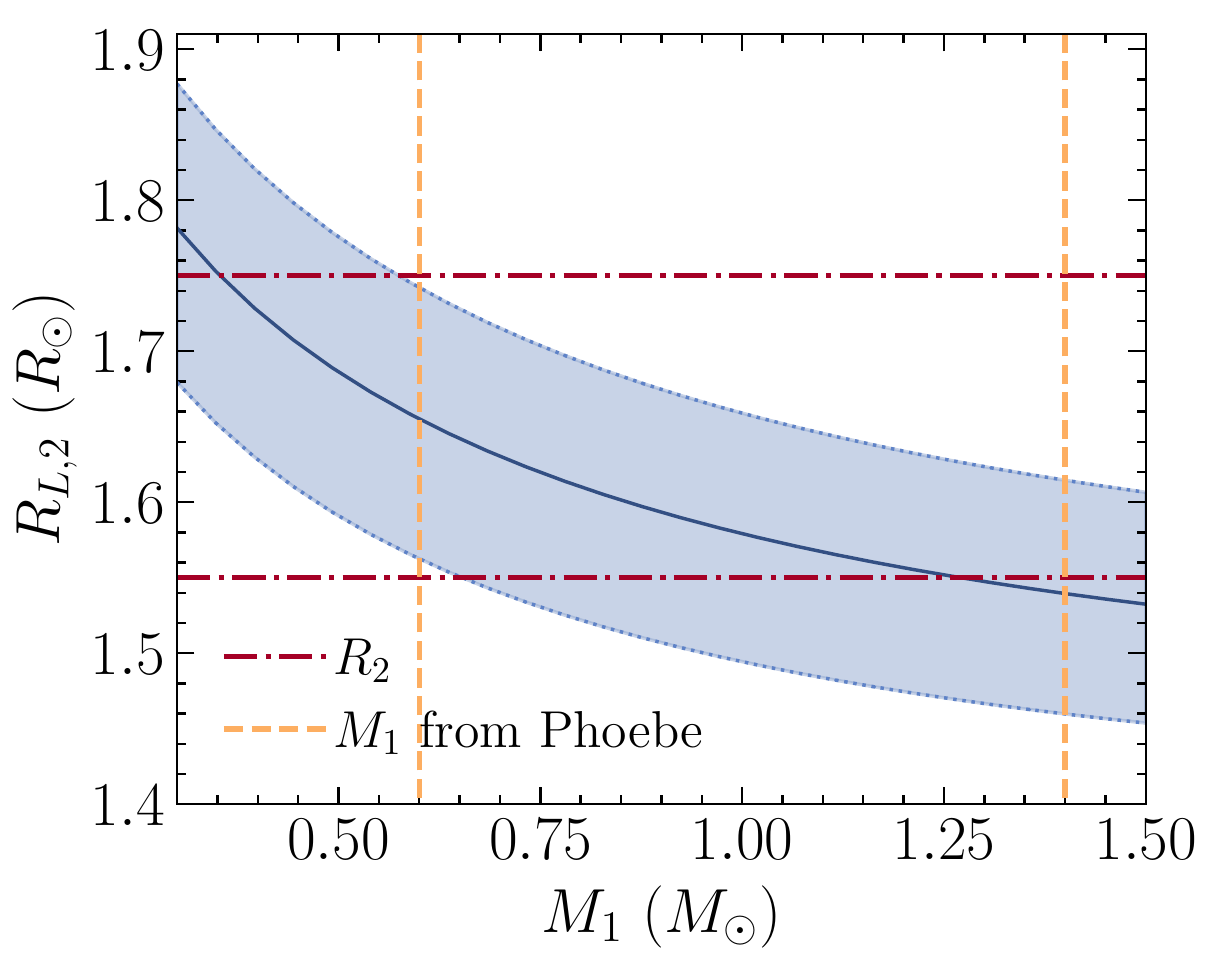}
    \caption{\textit{Left}: Phase-folded photometry of OGLE-BLG504.12.201843 during the quiescent state compared with two \textsc{Phoebe2} models. Data are shown in blue and two different models are overplotted. The blue line shows an early~F companion of 1.6$M_\odot$ with a WD of 1.4$M_\odot$ and $i=45$°, with gravity darkening and irradiation of the secondary. The red line shows an early~F companion of 1.6$M_\odot$ with a WD of 1.4$M_\odot$ and $i=60$°, with only irradiation of the secondary. In both models the primary has a temperature of $3\times 10^4$K. \textit{Right}: Estimation of the Roche lobe radius of the secondary $R_{L,2}$ using Equation \eqref{eq:rocheradius} with different primary masses $M_1$ and a secondary mass $M_2=1.6\pm 0.2 M_\odot$. The constraints on $M_1$ obtained with \textsc{Phoebe2} are shown as yellow vertical dashed lines and the range of possible secondary radii $R_2$ is indicated as red horizontal dashed lines.}
    \label{fig:constraints}
\end{figure*}

The growth of emission lines observed in the outburst spectra (see Figure~\ref{fig:HeII}) could be explained by an accretion disc: the lines grow in intensity as temperature increases in the optically thin parts of the disc. However, these lines are not double-peaked, which is generally expected from an accretion disc. This could be related to a moderate inclination of the system or some other physical process. For instance some Nova-like~CVs show single-peaked disc emission that are well reproduced by disc winds \citep{inight.2021b}. There are also high inclination Nova-like CVs, called SW~Sex systems \citep{dhillon.2013}, that only show double-peaked signals for a small portion of their orbital period.

Even though the accretion disc provides a crude explanation for the emission lines in outbursts, we see strong differences between the spectra of O-201843 and those of known systems with thermally unstable accretion discs. Efforts in identifying the spectroscopic signatures of DNe show that one should generally expect Balmer emission cores with absorption wings in quiescence and Balmer absorption lines in outburst. This behaviour seems to be explained by a change in optical thickness as temperature increases in the disc. Yet there seem to be a substantial amount of systems that deviate from this rule, showing no or barely perceptible emission cores in quiescence and some emission lines during the outburst \citep{Han.2020, Morales.2002}. Thus, the lack of emission lines in the quiescence spectra of O-201843 is concerning but does not invalidate the hypothesis of an outbursting disc.

The most straightforward explanation for the absence of emission cores is that they might be undetectable in our spectra. Given the low signal-to-noise ratio and resolution of the spectra, along with the brightness of the secondary, it is possible that small emission cores are hidden. This might also account for the shallowness of the stellar absorption lines. O-201843 could then have a bright secondary similarly to the unusual CV V1129~Cen, which shows faint 0.6-0.8~mag outbursts in the $V$~band with a recurrence timescale of roughly a year \citep{walter.2006}. This system does not show the usual emission lines caused by the accretion disc and \cite{bruch.2017} hypothesised that this system could be a DN with a bright type~F secondary star that outshines the disc. The issue with this hypothesis lies in the lack of change in absorption line profiles during the outburst, when the disc outshines the rest of the system. As stated before, one expects to detect absorption lines originating from the disc during the outburst. In the case of a bright secondary, these lines could blend with the stellar absorption lines, but it is unclear why we do not observe any significant change in the profile of these lines when the disc undergoes an outburst.

Another possible explanation for the lack of emission lines during quiescence is given by \cite{Idan.2010}. They modelled the spectra of cold accretion discs using the DIM and found an optically thick disc with Balmer absorption lines and no emission lines. They argue that optically thin regions in quiescence should only appear in the disc photosphere or in winds \citep{Matthews.2015}, which would only appear during outbursts or if there is significant irradiation of the disc by the WD. In the case of O-201843, the long timescales of the outburst seem to suggest that the disc is initially cold and the irradiation from the WD is minimal. So the quiescent spectra could be dominated by absorption lines, and consequently the Balmer lines we observe would be a blend of the lines from the disc and the secondary. As the temperature increases during the outburst, some regions of the disc might become optically thin and cause the emission lines.

Additionally, we tried to fit the outburst spectra taken by the MIKE spectrograph with \textsc{The Payne} spectral models \citep{Ting.2019} to see if we could extract some stellar parameters. No good fit was found because the spectra are too smooth and the few absorption lines that were detected are uncharacteristically shallow. This could mean that the profile of the absorption lines is indeed modified by the accretion disc, either with undetectable emission cores or blends of absorption lines.

Altogether, the spectra of O-201843 do not exclude the presence of an accretion disc despite not showing the expected features. Additional information could be obtained by examining the evolution of the continuum during the outburst, however, as stated in Section~\ref{sec:spectranalysis}, this is not possible with the current data. It is therefore important to get more spectra of the system during outbursts.

\subsection{Geometry of the system}\label{sec:geom}
The constraints on the nature of the secondary obtained with the spectra of O-201843 require an unusually bright companion for a CV. It is thus necessary to check whether these configurations are consistent with the known parameters of the system.

\begin{figure*}
    \includegraphics[width=\linewidth]{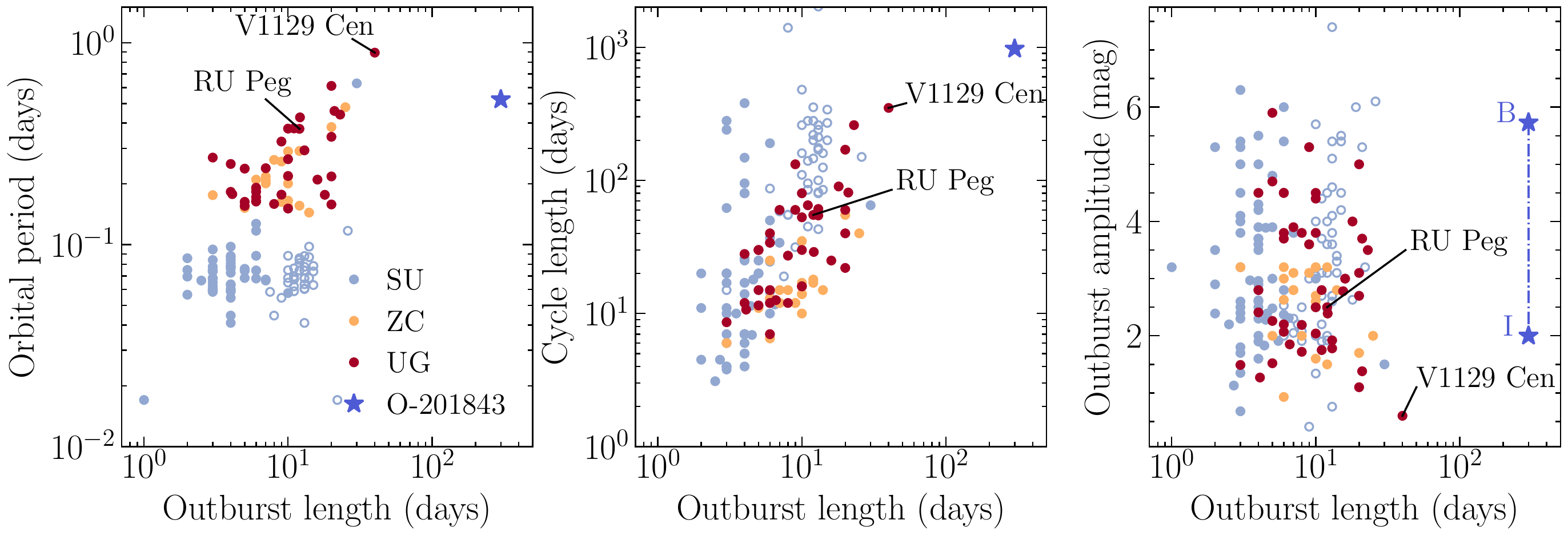}
    \caption{Comparison of the properties of OGLE-BLG504.12.201843 with the catalog of DNe compiled by \protect\cite{otulakowska.2016}. The parameters of O-201843 are shown with a blue star, the catalog in DNe is separated in SU~UMa (blue), Z~Cam (yellow) and U~Gem (red) systems. We also indicate the properties of V1129~Cen and RU~Peg. Left panel: orbital period vs outburst length. Middle panel: average cycle length vs outburst length. Right panel: outburst $V$~amplitude vs outburst length, the outburst amplitude of O-201843 coming from DASCH (in the $B$~band) is denoted with "B" and the one coming from OGLE (in the $I$~band) is denoted with "I". For SU~UMa systems, the parameters of regular cycles are denoted with full points while circles indicate the properties of supercycles.}
    \label{fig:comparison}
\end{figure*}

First, we use the orbital variability in Figure~\ref{fig:photometryphase} to estimate some of the properties of the binary, specifically the mass~ratio~$q$ and inclination~$i$. We try to reproduce the ellipsoidal shape of the quiescent light curve using \textsc{Phoebe2} \citep{Prsa.2016}. We set up the system as a semi-detached binary with a period of 0.523419~days with an early~F secondary star ($M_2=1.6M_\odot$). We vary the mass of the WD primary $M_1$ and inclination of the system $i$, and set the radius of the WD to follow the relation $R\propto M^{-1/3}$. We run the models with either gravity darkening, secondary irradiation or both to see which combination is able to reproduce the orbital variability.

We find that secondary irradiation alone fails to induce a difference in the two minima observed in the quiescent variability of the system for primary temperature below $3\times10^4$~K. The effect of irradiation is still small at higher temperatures, which is likely due to the large separation between the secondary and the primary. The insignificance of irradiation by the primary means that the observations are consistent with a relatively cold primary. Conversely, gravity darkening manages to replicate both the shape and minima difference of the variability. The only difference we observe when changing $M_2$ is a lower $i$ for a higher $M_2$, which is likely due of the degeneracy between $i$ and mass ratio $q=\frac{M_2}{M_1}$. For an early~F star, the most successful models have $M_1$ between 0.7 and 1.4~$M_\odot$ ($1.29\lesssim q \lesssim2.57$) with $i$ between 45° and 50°. We show an example of a successful model in the left panel of Figure~\ref{fig:constraints} for the case of $M_2=1.6M_\odot$, as well as an example of the contribution of secondary irradiation. The moderate inclination is consistent with the lack of eclipse in quiescence, and might contribute to the lack of double-peaked signals in the disc spectra.

An early~F secondary is unusually young and large for a CV. We check how such a secondary fits within the orbit parameters by calculating the semi-major axis $a$ of the system:
\begin{equation}\label{eq:kepler}
    a^3=\frac{GMP_\textrm{orb}^2}{4\pi^2},
\end{equation}
where $M=M_1+M_2$. Under the early~F secondary assumption ($M_2\simeq1.6M_\odot$), this equation yields $a=3.7 R_\odot$ for $M_1=0.7M_\odot$ and $a=3.9 R_\odot$ for $M_1=1.4M_\odot$. Thus an early~F secondary with $R_2\simeq1.7R_{\odot}$ would fit within the system orbit. A study of population synthesis of CVs by \cite{Goliasch.2015} shows that CVs can form with companions of spectral type earlier than~K. Using MIST isochrones \citep{Dotter.2016, Choi.2016}, we estimate that the maximum main-sequence age of a $1.6M_\odot$ star is around  $1.8\times10^9$~years. According to the WD cooling tracks from \citep{Fontaine.2001} which leaves enough time for a WD primary to cool to temperatures as low as $10000$~K. We note however that accretion onto the WD primary should increase its temperature. Therefore, depending on the mass transfer rate of the system, the WD temperature might significantly diverge from the low temperatures mentioned above.

Overall, an early~F secondary seems to fit within the observed parameters of the system. The next step is to establish whether such a companion is consistent with the observed outbursts, especially the long timescales and the corresponding mass transfer rates.

\subsection{Mass transfer} \label{sec:masstransfer}
If the outbursts of O-201843 originate from a thermally unstable accretion disc, then the mass transfer from the secondary and in the disc should fulfill certain criteria.

Firstly, an accretion disc requires that the secondary overfills its Roche lobe. Using the Roche lobe radius approximation from \cite{Eggleton.1983}:
\begin{equation}\label{eq:rocheradius}
    \frac{R_L}{a}=\frac{0.49q^{2/3}}{0.6q^{2/3}+\ln(1+q^{1/3})},
\end{equation}
together with Equation~\eqref{eq:kepler} allows to estimate the Roche lobe radius of the secondary $R_{L,2}$ given a specific mass ratio. We evaluate $R_{L,2}$ over all possible primary masses in the cases of an early F-type main sequence secondary with mass $M_2=1.6\pm 0.2 M_{\odot}$. In the right panel of Figure~\ref{fig:constraints}, we compare $R_{L,2}$ to the range of possible secondary radii $R_2=1.6\pm 0.15 R_{\odot}$ for an early~F companion. We also indicate the constraints on the primary mass $M_1$ obtained with \textsc{Phoebe2}. According to these estimations, it seems possible that an early~F companion overflows its Roche lobe. We checked these estimations for cases of smaller secondary stars and found that in this system, a main sequence secondary of type later than late type~F cannot overfill its Roche lobe. This is to be expected since the dimensions of the system are quite large for a CV, and thermally unstable accretion discs seem to mainly occur in binaries with much closer orbits and smaller secondary stars. We therefore conclude that in the case of a main sequence companion, the presence of an accretion disc requires the secondary star to be of type no later than~F. We note however that in the case of a more evolved secondary, the requirement derived above does not hold anymore.

Secondly, the recurrent outbursts in the disc should be explained by the DIM, in which the stability of the disc depends on the local mass transfer rate. \cite{Kalomeni.2016} studied the evolution tracks of CVs, in particular they included the treatment of unstable accretion discs using the critical mass transfer rates inferred by the DIM. Their results show that for the orbital period of O-201843, an early~F companion allows for the formation of thermal-viscous instabilities in the disc. However, they also show that the average mass transfer rate $\langle\dot{M}\rangle$ of such a system would be above $10^{-8.8} \textrm{M}_\odot/\textrm{yr}$, meaning the system would fall in the regime of super-soft X-ray sources. In this case, the additional X-ray radiation may heat up the disc sufficiently to prevent the development of thermal instabilities. Furthermore, according to \cite{Townsley.2009}, a system with $\langle\dot{M}\rangle>10^{-8}\textrm{M}_\odot/\textrm{yr}$ results in  a WD primary temperature $T_1>25000$~K for $M_1=0.6M_\odot$ or $T_1>50000$~K for $M_1=0.9M_\odot$, which is inconsistent with the idea of a cool WD primary and disc. It is therefore unclear whether thermal-viscous instabilities actually occur in O-201843.

\subsection{Comparison to DNe}\label{sec:comparison}
\begin{figure}
    \includegraphics[width=\columnwidth]{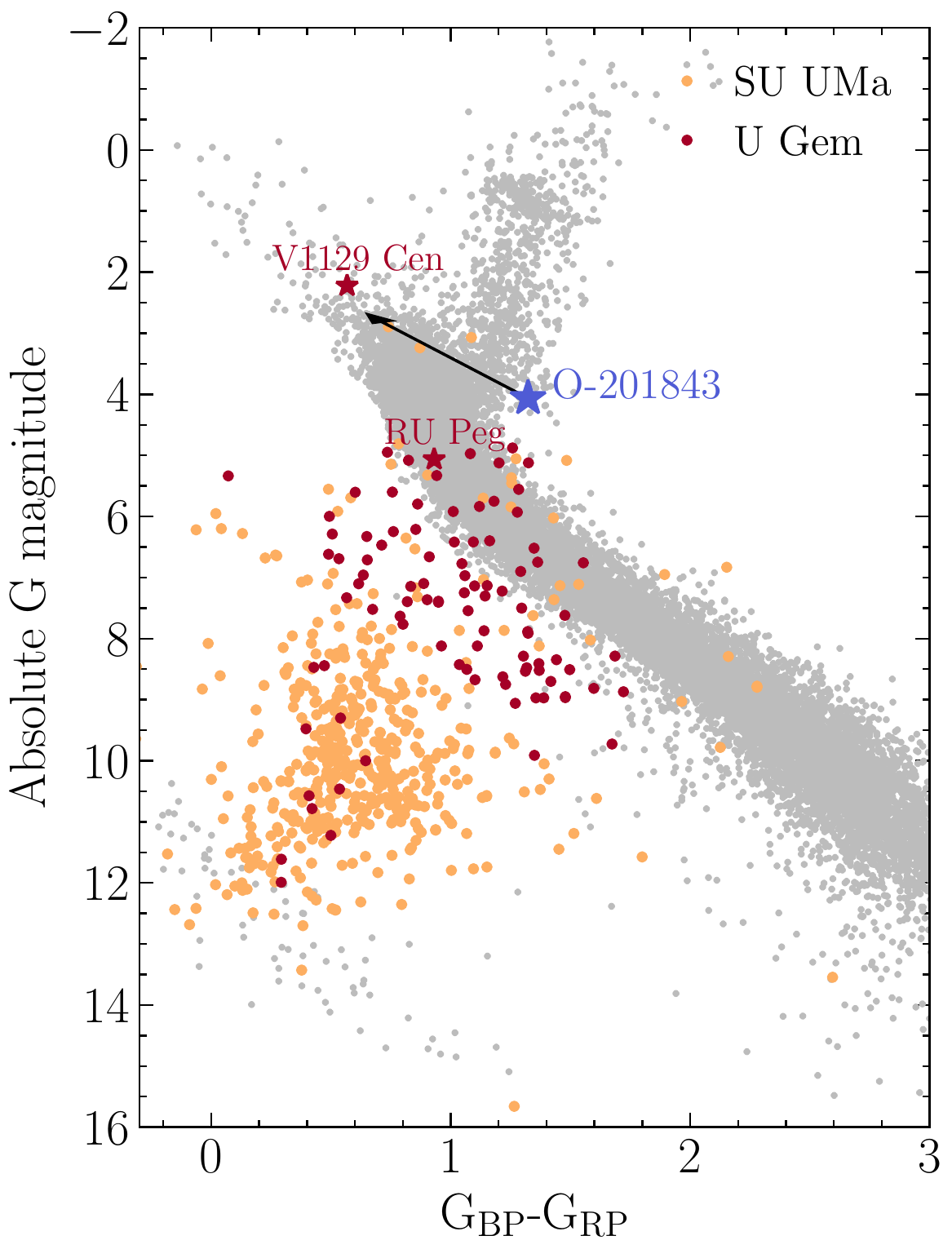}
    \caption{OGLE-BLG504.12.201843 in the HR diagram with a catalog of DNe obtained by merging the ``Gold'' sample of CVs from \protect\cite{inight.2021a} with the CV sample from \protect\cite{Abril.2020}. O-201843 is denoted with a blue star and DNe are separated in SU~UMa (yellow) and U~Gem (red) systems. The black arrow shows the effect of a reddening of $E(B-V)=0.5$~mag. Points in grey are data selected from the Gaia Early Data Release 3 (EDR3, \citealt{gaia.2021}) by requiring that the sources have parallax errors lower than 10\% of their parallax. This criterion allows to compute the absolute magnitudes with the distance as inverse of the parallax.}
    \label{fig:HR}
\end{figure}

Overall, the regular recurrence and shape of the outbursts of O-201843, along with their asymmetry, are reminiscent of long outside-in outbursts of U~Gem type DNe: a fast rise followed by a decline that is slow at first and then accelerates. While the long timescale of the outbursts is analogous to superoutbursts \citep{Rivera.2020}, \cite{Mroz.2016} did not find any secondary luminosity modulations that would correspond to superhumps. Besides, the shape of the outbursts does not resemble those of superoutbursts, they are more similar to normal DNe outbursts despite their long timescales. We conclude this study by comparing the photometry O-201843 to that of other known DNe in hope that it will help to draw a conclusion on the nature of the system.

An obvious divergence from DNe occurs during the quiescent state, which shows a slow 0.75~mag rise in brightness. Despite being a feature predicted by the DIM, we could not find any mention of the observation of such brightenings in the literature, although the SU~UMa type DN V363~Lyr does show a small brightening before both outbursts and superoutburst \citep{kato.2021}. Interestingly, \cite{smak.2000} notes that this transitional state is one of the weaknesses of the DIM since it has never been observed in other DNe despite being essential to occurrence of the outburst. This could mean some additional mechanism is involved in DNe outbursts that does not operate in O-201843. The lack of such mechanism could be linked to the unusually long recurrence time of the outbursts of O-201843.

Additionally, we do no find records of the small recurring flares in quiescence mentioned in Section~\ref{sec:flares} in other DNe. They could be much more short-lived and harder to observe in the case of normal outburst cycle time. However consecutive flares were observed after superoutbursts in WZ~Sge systems \citep{kato.2015}, which were successfully reproduced by \cite{hameury.2021} by adding several mechanisms to the DIM, such as variation of the mass transfer from the secondary, disc truncation and irradiation of the disc. In the case of O-201843, these flares could be small instabilities that fail to propagate far in the disc. Their symmetric shape are reminiscent of inside-out outbursts, which are triggered close to the inner edge of the disc and involve a much smaller portion of the disc than outside-in outbursts. Overall these flares are a very interesting feature and are most likely linked to disc instabilities.

In Figure~\ref{fig:comparison}, we compare the properties of O-201843 to the catalog of DNe compiled by \cite{otulakowska.2016}. Here, the cycle length should be interpreted carefully because the time elapsed between cycles can vary a lot for one DN. Since we do not have $V$~band data for O-201843, we mark the outburst amplitude with a range set by the photographic amplitude from DASCH and $I$~band data from OGLE. Additionally, some DNe in this catalog also display superoutbursts (SU~UMa systems) and in this case we differentiate the parameters of cycles and supercycles with different symbols. The cycle length is the time elapsed between two normal outbursts and the length of supercycle is the time elapsed between two superoutbursts. The recurrence time of O-201843 is unusually long compared to that of DNe, where outbursts occur every few weeks or months. The length of the outburst itself is also generally much shorter, around a couple of days or weeks instead of almost a year in O-201843. From both left and middle panels in Figure~\ref{fig:comparison} we can clearly see that O-201843 is an outlier. As shown in the middle panel, \cite{otulakowska.2016} suggest there could exist a relation between the outburst length and the cycle length. In this case, O-201843 would actually follow this relation despite having such extreme properties. The right panel shows that the amplitude of the outburst is relatively normal.

In Figure~\ref{fig:HR}, we show the position of O-201843 in the Hertzprung-Russel~(HR) diagram alongside a catalog of DNe obtained by merging the "Gold" sample of CVs from \cite{inight.2021a} with the CVs sample from \cite{Abril.2020}. Both samples were selected from Gaia~DR2 \citep{gaia.2018} by considering objects with accurate parallax that were cross-matched with various CVs catalogues. We plot the HR diagram using similar criteria and compute distances as inverse parallax, which gives 2.35~$\pm$~0.29~kpc for O-201843. We estimated the reddening $E(B-V)$ of O-201843 by measuring equivalent widths of Na~I D lines in our high resolution spectra and using the relations of \cite{poznanski.2012}. We obtained values of $E(B-V)$ between 0.1 and 0.45~mag and chose to indicate the effect of $E(B-V)=$~0.5~mag in the HR diagram. From Figure~\ref{fig:HR} we can see that the WD or accretion disc is dominant in most DNe. Only a couple of systems (mainly U~Gem DNe) and O-201843 appear closer to the main sequence, meaning they are likely dominated by a main sequence secondary star. We also indicated the properties of two systems that are quite close to O-201843 in Figure~\ref{fig:comparison}, namely RU~Peg and V1129~Cen. While RU~Peg is quite ordinary, V1129~Cen has more extreme characteristics which were discussed in Section~\ref{sec:accdisc}. Nonetheless, V1129~Cen has much shorter outburst and cycle length than O-201843. The similar luminosity and color between O-201843 and V1129~Cen gives further support to our hypothesis of a bright early~F secondary star. We note that reddening corrections would make the secondary even brighter and bluer, as indicated by the arrow in Figure~\ref{fig:HR}.

Altogether, the photometry of O-201843 shows strong similarities with DNe but also important differences. While these peculiarities have not been observed in other DNe, it is possible that O-201843 is an extreme U~Gem DN, or at least is somehow linked to DNe.

\section{Conclusion}\label{sec:conclusion}

To summarize, we have analysed optical photometry and spectroscopy of O-201843 and suggest that O-201843 is a U~Gem type DN with extreme properties. The photometry shows clear features of an accretion disc from which the outbursts can originate. It also display two unusual features that, to our knowledge, are only rarely observed in other DNe. The slow brightening preceding the outbursts is a prediction of the DIM and its absence in regular DNe could indicate that some physical process usually suppresses it. The small flares with amplitude $\lesssim 0.2$~mag in $I$~band detected during quiescence are another peculiar feature. We suggest that they might be small outbursts that fail to propagate far in the disc.

The analysis of the spectroscopy shows Balmer absorption that can either come from an early~F secondary or the accretion disc. We however do not see the features of a WD and the usual features of an unstable accretion disc, i.e. no emission lines or cores in quiescence and no additional Balmer absorption in outburst. A possible explanation is that the disc is not or weakly irradiated by the primary WD, rendering it unusually cold. Additionally, the emission lines appearing during the outburst lack the double-peaked signal that one would expect from an accretion disc. As suggested by the orbital variability of the system, the inclination might just be low enough for the lines to be single-peaked. Obtaining time-resolved spectroscopy of this system could help to understand the lack of double-peaked lines as well as accurately constrain some of the parameters of the system such as its mass ratio and inclination. It could also provide insights on how the system changes during the small flares appearing during quiescence. Additionally, such spectra would permit a similar analysis to what what was done by \cite{kara.2021}, who used Doppler tomography \citep{marsh.1988} on the disc emission lines to obtain information on the velocity structure of the disc. Moreover, new spectra, especially in UV, could also help elucidate the nature of the primary and secondary.

Finally, the outburst duration and cycle length would be unprecedented among other U~Gem-type DNe. One potentially similar object to O-201843 is V1129~Cen, which has similar orbital period and location in the HR diagram, but its outburst and cycle length and amplitude are still significantly smaller than those of O-201843. If O-201843 is not a DN, it could be a new type of CVs that is tightly linked to DNe. Further study of this system should help understanding thermal-viscous instabilities and outbursts in accretion discs. Furthermore, the very long timescales of the system allows to see features that might not be possible to observe otherwise (e.g. flares) and it will be of great benefit to study them. The next outburst of O-201843 should begin in May 2023, peak around July 2023 and end in February 2024. The end of this outburst will not be visible due to conjunction with the Sun. The following outburst, starting in January 2026 and ending in October 2026 with a peak around March 2026, should be fully visible.

\section*{Acknowledgements}

We thank Juna Kollmeier, Julio Chanamé and Doron Kushnir for contributing to the spectroscopic observations of OGLE-BLG504.12.201843 and Krzysztof Stanek for coordinating these observations. CL thanks Yuan-Sen Ting for his valuable insights and his help with \textsc{The Payne}. We thank the referee for their helpful comments.

The work of CL and OP has been supported by INTER-EXCELLENCE grant LTAUSA18093 from the Ministry of Education, Youth, and Sports. The research of OP has been supported also by Horizon 2020 ERC Starting Grant ‘Cat-In-hAT’ (grant agreement no. 803158). 
MP is supported by the SONATINA grant 2020/36/C/ST9/00103 from the Polish National Science Center. 
Support for JLP is provided in part by ANID through the Fondecyt regular grant 1191038 and through the Millennium Science Initiative grant ICN12\textunderscore009, awarded to The Millennium Institute of Astrophysics, MAS. 
JS was supported by the Packard Foundation.

We thank the support from Chilean Time Allocation Committee (CNTAC) through program ID CN2017B-85.
The DASCH project at Harvard is grateful for partial support from NSF grants AST-0407380, AST-0909073, and AST-1313370.

\section*{Data Availability}
OGLE data are available on a reasonable request to A. Udalski. DASCH data are publicly available. Spectroscopic data are available on a reasonable request to C. Landri.
 



\bibliographystyle{mnras}
\bibliography{ogle-bin} 





\appendix

\onecolumn

\section{Spectra}
\label{sec:spe}

\begin{figure*}
    \includegraphics[width=0.9\textwidth]{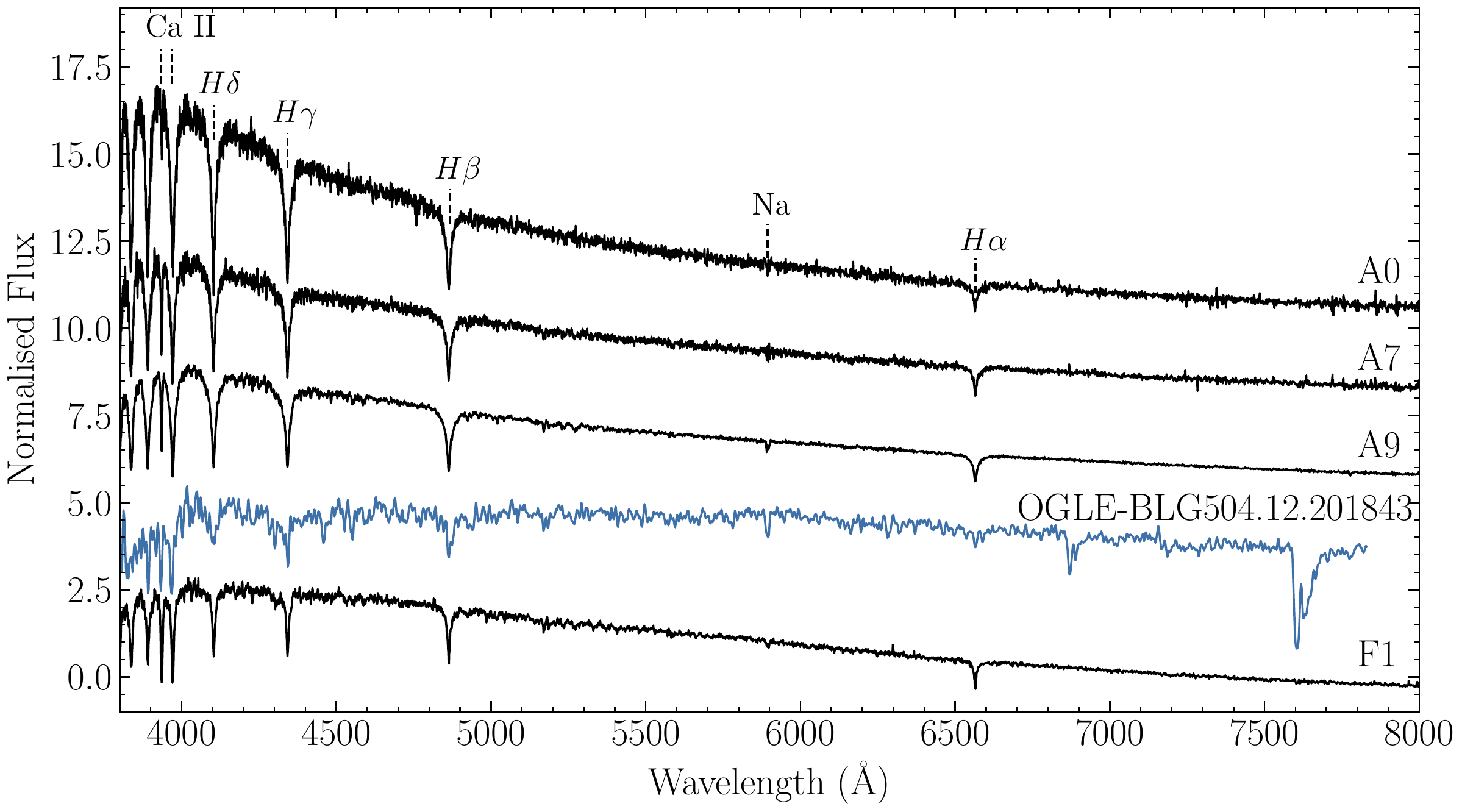}
    \caption{Full Spectra of O-201843 in quiescence (taken on 2017-10-24) with identification of interesting lines and template spectra from \textsc{PyHammer} \citep{Kesseli.2017, Roulston.2020}. Quantitative parameters of the lines are given in Table~\ref{tab:spectra}.}
    \label{fig:fullspectra}
\end{figure*}

\begin{figure*}
    \includegraphics[width=0.4\columnwidth]{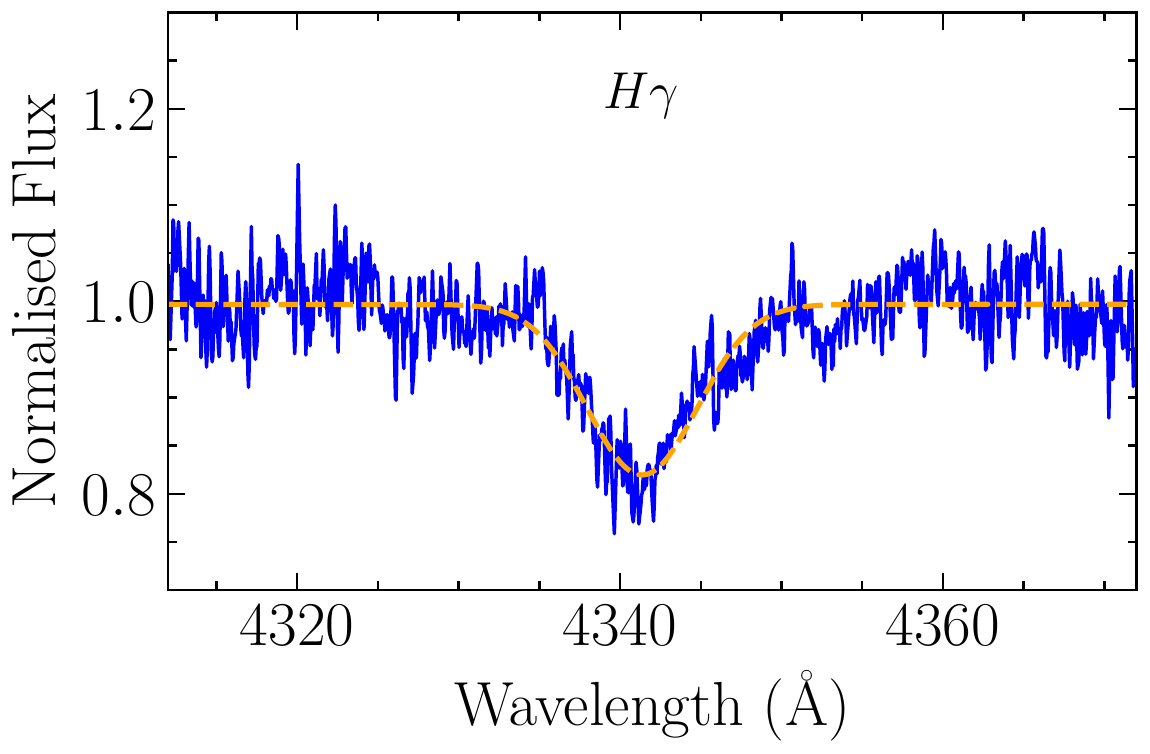}
    \includegraphics[width=0.4\columnwidth]{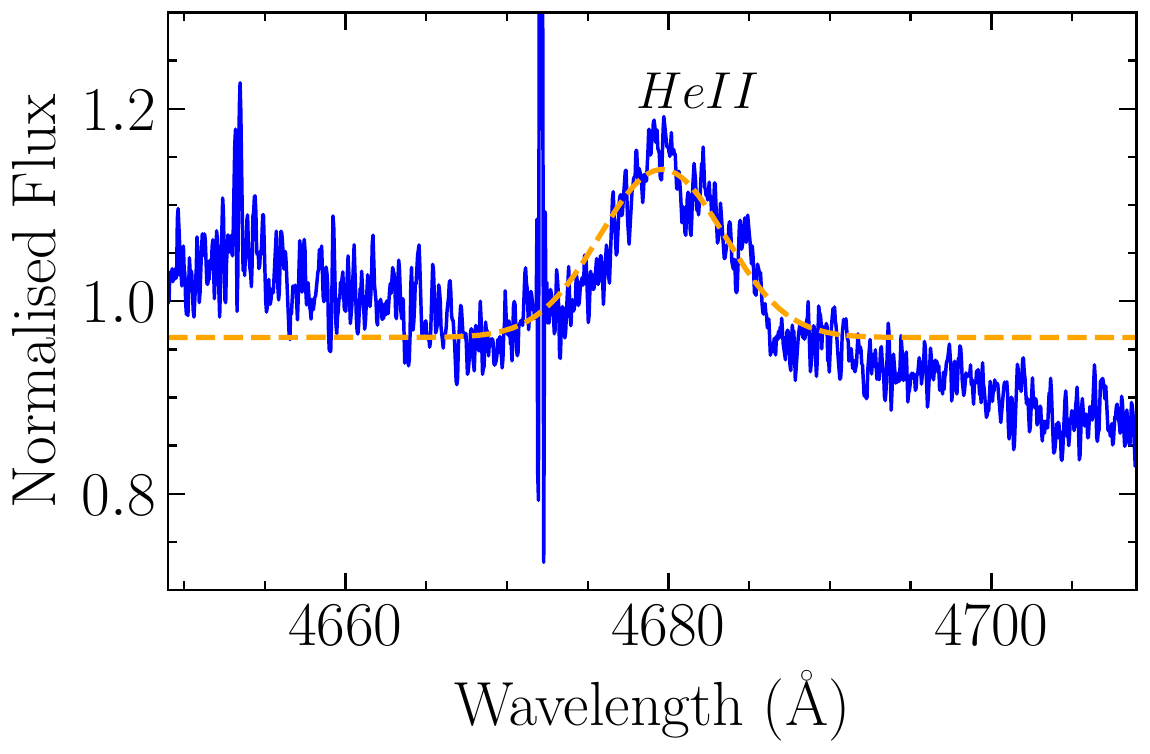}
    \includegraphics[width=0.4\columnwidth]{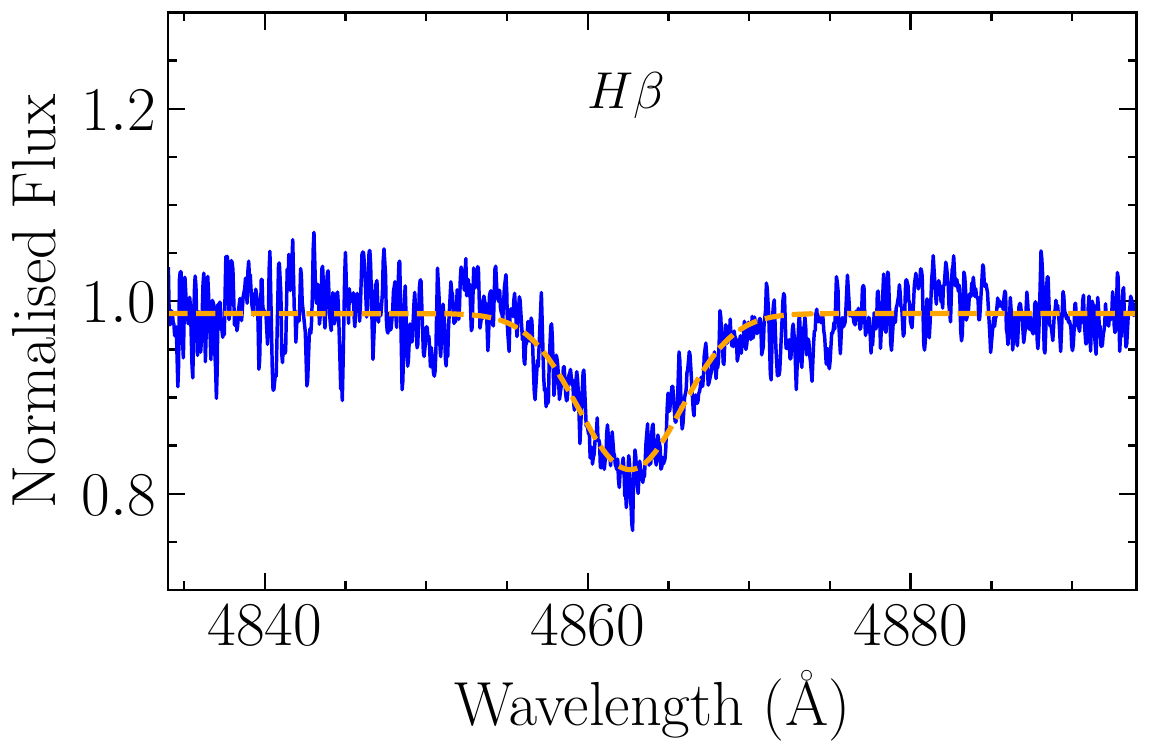}
    \caption{Portions of the spectrum of O-201843 during an outburst (taken on 2018-03-06) with identification of interesting lines. Quantitative parameters of the lines are given in Table~\ref{tab:spectra}.}
    \label{fig:spectra}
\end{figure*}

In Figure~\ref{fig:fullspectra}, we show the full spectrum of O-201843 during quiescence, indicate the interesting lines and compare it to the spectra of typical A0, G2 and F2 stars. In Figure~\ref{fig:spectra}, we show wider portions of our spectra and identify of interesting lines.


\bsp	
\label{lastpage}
\end{document}